\documentclass{jfm}
\usepackage{graphicx}
\usepackage{natbib}
\usepackage{tikz}
\usepackage{gensymb}
\usepackage{amsmath}
\usepackage{wrapfig}

\newcommand{\beq}{\begin{equation}}  
\newcommand{\eeq}{\end{equation}}  
\newcommand{\bea}{\begin{align}}  
\newcommand{\eea}{\end{align}}  
\newcommand{\Pra}{\textrm{Pr}}  
\newcommand{\Ray}{\textrm{Ra}}  
\newcommand{\Nus}{\textrm{Nu}}  
\newcommand{\ReyS}{\textrm{Re}_S}
\newcommand{\Rom}{\textrm{R}_\Omega}  
  
\newcommand{\hatt}{}

\title[2D Rayleigh-B\'enard and Rotating Plane Couette flows]{Exact relations between
Rayleigh-B\'enard and rotating plane Couette flow in 2D}

\author[B. Eckhardt, C.R. Doering, and J. P. Whitehead]%
{Bruno Eckhardt$^{1}$,  Charles R. Doering$^{2}$, and Jared P. Whitehead$^{3}$}

\affiliation{$^1$ Fachbereich Physik, Philipps-Universit\"at Marburg, D-35032 Marburg, Germany
\\[\affilskip]
$^2$ Center for the Study of Complex Systems, Department of Mathematics and \\
Department of Physics, University of Michigan, Ann Arbor, MI 48109 USA\\[\affilskip]
$^3$ Department of Mathematics, Brigham Young University, Provo, UT 84602, USA
}
\pubyear{2020}
\volume{--}
\pagerange{--}

\begin{document}

\maketitle

\begin{abstract}
Rayleigh-B\'enard convection (RBC) and Taylor-Couette Flow (TCF) are two paradigmatic fluid dynamical systems 
frequently discussed together because of their many similarities despite their different geometries and forcing.
Often these analogies require approximations, but in the limit of large radii where TCF becomes rotating plane Couette flow (RPC) exact relations can be established.  When the flows are restricted to two spatial degrees of freedom there is an exact specification that maps the three velocity components in RPC to the two velocity components and one temperature field in RBC.
Using this, we deduce several relations between both flows:
(i) The Rayleigh number $\Ray$ in convection and the Reynolds $\ReyS$ and rotation $\Rom$ 
number in RPC flow are related by $\Ray= \ReyS^2 \Rom (1-\Rom)$.  
(ii) Heat and angular momentum transport differ by $(1-\Rom)$, explaining why angular momentum transport is not symmetric around $\Rom=1/2$ even though the relation between $\Ray$ and $\Rom$ has this
symmetry.  This relationship leads to a predicted value of $\Rom$ that maximizes the angular momentum transport that agrees remarkably well with existing numerical simulations of the full 3D system.
(iii) One variable in both flows satisfy a maximum principle i.e., the fields' extrema occur at the walls.
Accordingly, backflow events in shear flow \emph{cannot} occur in this two-dimensional setting.
(iv) For free slip boundary conditions  on the axial and radial velocity components, previous rigorous analysis for RBC implies that the azimuthal momentum transport in RPC is bounded from above by $\ReyS^{5/6}$ with a scaling exponent smaller than the anticipated $\ReyS^1$.
\end{abstract}
 


\section{Introduction}
Rayleigh-B\'enard convection (RBC), the buoyancy-driven motion of a fluid heated from below, and Taylor-Couette Flow (TCF) wherein a fluid is sheared between two rigid co-rotating cylinders,
are paradigms in the physical and engineering sciences and have been studied extensively to gain insights into turbulence.
It has long been recognized that despite their qualitative differences they share many features, both physically and mathematically. 
Indeed, the comparison between RBC and TCF goes back nearly to the original definition of these canonical fluids problems.
As stated in \cite{Je1928}:
\begin{quote}
\emph{Prof. G. I. Taylor and Major A. R. Low have both suggested to me that there should be an analogy between the conditions in a layer of liquid heated below and in a liquid between two coaxial cylinders rotating at different rates.}
\end{quote}
\cite{Je1928} considered this analogy in the context of linear stability of the basic states (pure conduction for RBC and axisymmetric laminar flow for TCF), a line of reasoning quantified further by \cite{Ch1961}.
Subsequent investigations into the onset of convective and shear turbulence have led to significant advances in pattern formation, the mathematical theory of chaotic and nonlinear dynamics, and insight into the influence and interaction of linear and nonlinear instabilities \citep{GS1975,AB1978,manneville2010instabilities,chossat2012couette}.

This analogy was first extended to turbulent flow by \cite{Br1969}, with further contributions due to \cite{DuHe2002} and \cite{EcGrLo2007a,EcGrLo2007b}.  The basis for these analogies is the identification and comparison of corresponding quantities between the two systems, such as the total dissipation and global transport of physically motivated quantities like heat or angular momentum.  The similarities between RBC and TCF then lead to relations between the non-dimensional parameters of the system and allow for direct comparisons of the pertinent physical quantities.
As noted by \cite{BrEcSc2017} and demonstrated in direct numerical simulations, the similarity between TCF and RBC gives rise to similar behavior not only in the mean properties but also the fluctuations indicating that an even more precise comparison may be possible.

Correspondence between RBC and TCF---more specifically Plane Couette Flow (PCF)---also extends into the realm of rigorous mathematical analysis.  Energy stability of the conductive state in RBC corresponds precisely (up to the appropriate change of variables) to energy stability of the laminar plane-parallel solutions of PCF.  As an extension of energy stability, the original upper bound analysis for statistically stationary heat transport in RBC introduced by \cite{Howard1963} transfers directly via relabeling and rescaling of variables to an upper bound analysis for the energy dissipation rate in PCF \citep{Busse1969,Howard1972}.
The subsequently developed background method for producing upper bounds \citep{DoCo1992} shares the same exact correspondence \citep{DoCo1994,DoCo1996,PlKe2003}.
Both Howard's approach and the background method are easily adapted to the cylindrical setting of TCF \citep{Nickerson1969,Constantin1994}.

There are profound differences between these two canonical problems as well.  In particular RBC has a parameter with no correspondence in the TCF or PCF setting, namely the Prandtl number.
The dynamics and analysis of RBC in the large Prandtl number limit---see, e.g., \cite{DoOtRe2006,OtSe2011,WhDo2012}---has no counterpart in TCF or PCF.
Physically relevant boundary conditions are also uniquely identified between convection and shear-driven flows.  For the specific case of two-dimensional RBC between free-slip isothermal boundaries, for example, \cite{WhDo2011a} showed that the convective heat flux at arbitrary Prandtl number is bounded above by the Rayleigh number to the $5/12$ power ruling out the conjectured `ultimate' $\sim$$\Ray^{1/2}$ scaling in this setting.
We can identify a corresponding situation in RPC where the azimuthal momentum transport is bounded above by the Reynolds number to the (perhaps unexpected) $5/6$ power.

The relation between RBC and TCF is complicated by the fact that the rotation in TCF flow does
not have a corresponding analogy in RBC. The comparison between the two flows shown
in \cite{BrEcSc2017} was therefore based on correspondences in mean transport. Here, 
we discuss consequences of an exact relation between 2D RBC and azimuthally symmetric 
TCF in the limit of a large cylindrical radius where it 
becomes rotating plane Couette flow (RPC) \citep{Nagata:1986wc,
Faisst:2000ti,Nagata:2013jn}.  
We emphasize at this point that for the geometry and restrictions described below, the resulting relations are exact, and without approximation.  That is to say, the 2D RBC system has an exact analogue in 2D RPC, the 2D TCF system in the limit of large radii.

\section{Derivation of the relations}
Exact relations between TCF and RBC are only possible if the spatial variables are restricted to two dimensions, as otherwise the number of dependent variables does not match.  In the full 3D setting RBC has three velocity components and one temperature, while TCF has only 3 components of velocity (both systems also have a pressure gradient).  Fully 2D RBC has two components of velocity and a temperature, while TCF or RPC that is independent of the azimuthal spatial coordinate will have three components of velocity, with the azimuthal velocity playing the role of the temperature in RBC. Some historical context for the analogy between RBC and TCF as well as comments on how it fails in 3D are explained in \cite{Veronis:1970dh}. 
The comparison between the two systems is relatively straightforward (the precise formulation of the analogy is given in an appendix to 
\cite{Nagata:2013jn}), but since the usual choice of coordinates does not provide an 
immediate translation between the two cases,
the main task in the following derivation is to keep track of the transformations 
in the dependent and independent variables. 

\subsection{2D Rayleigh-B\'enard}

For the 2D Rayleigh-B\'enard system there are two velocity fields and a temperature field.  
We take $\hatt{x}_1$ as the spanwise (horizontal) direction and impose periodicity in this direction,
$\hatt{x}_2$ is the neutral direction that is absent in the 2-d case, and $\hatt{x}_3$ points in the
direction of gravity
The velocity field then has components 
$\hatt{{\bf u}}=(\hatt{u}(\hatt{x}_1,\hatt{x}_3), \hatt{w}(\hatt{x}_1,\hatt{x}_3))$ that are restricted to be incompressible, $\hatt{\partial}_1 \hatt{u} + \hatt{\partial}_3 \hatt{w}=0$.
The boundary conditions on the velocity field are usually rigid 
$\hatt{u}=\hatt{w}=0$, or
free-slip (${\hatt{w}=0}$ and $\hatt{\partial}_3 \hatt{u} = 0$) at the top and bottom plates. The derivations in this Section 
do not depend on the specific boundary conditions on $\hatt{{\bf u}}$, however, in Section \ref{sect:freeslip} we will consider free-slip boundary
conditions specifically and the implications on maximal transport.

\begin{figure}
\begin{center}
$\begin{array}{c@{\hspace{.1in}}c} \multicolumn{1}{l}{} & \multicolumn{1}{l}{} \\
\scalebox{0.38}{\includegraphics{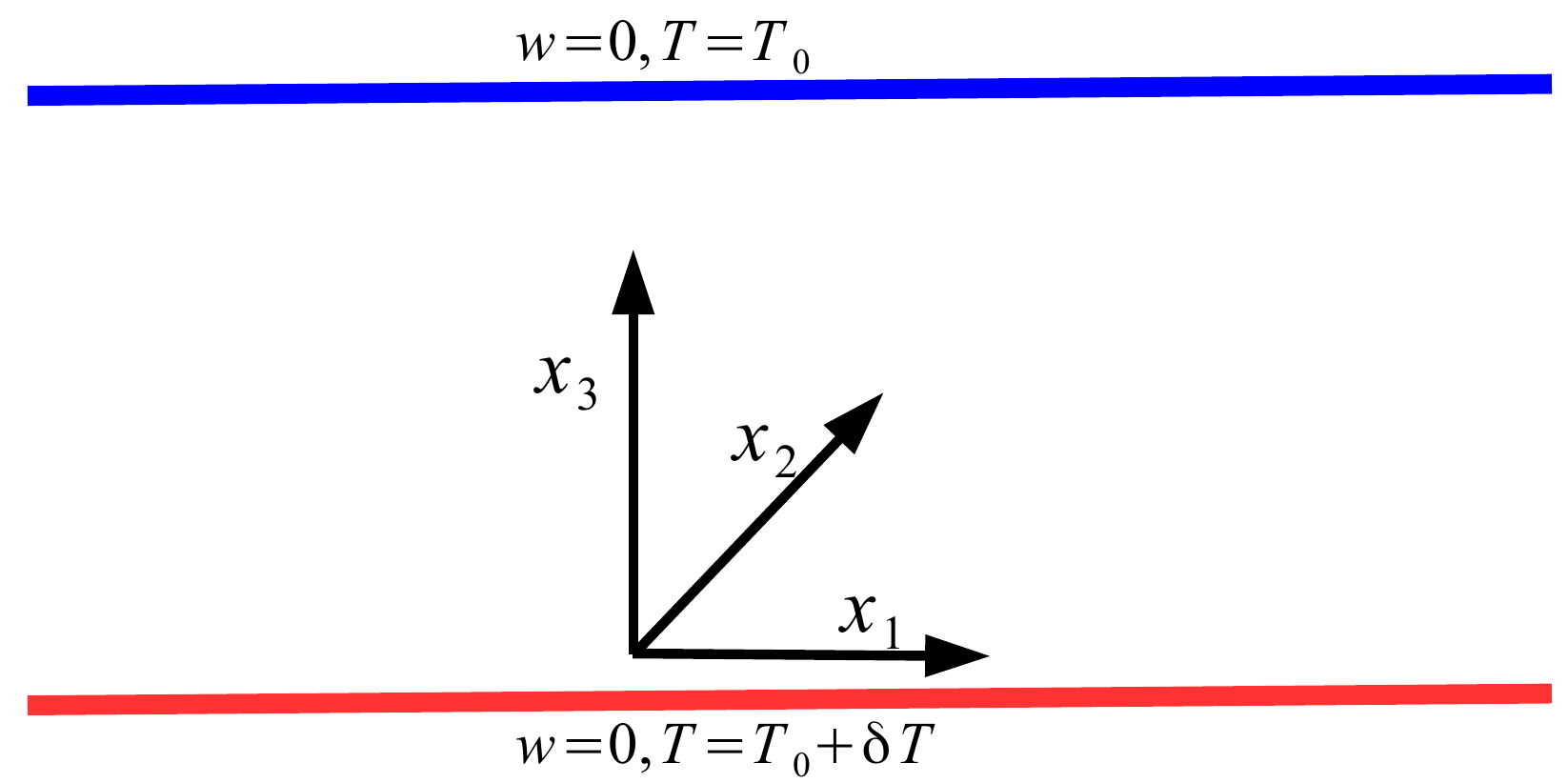}} &
\scalebox{0.3}{\includegraphics{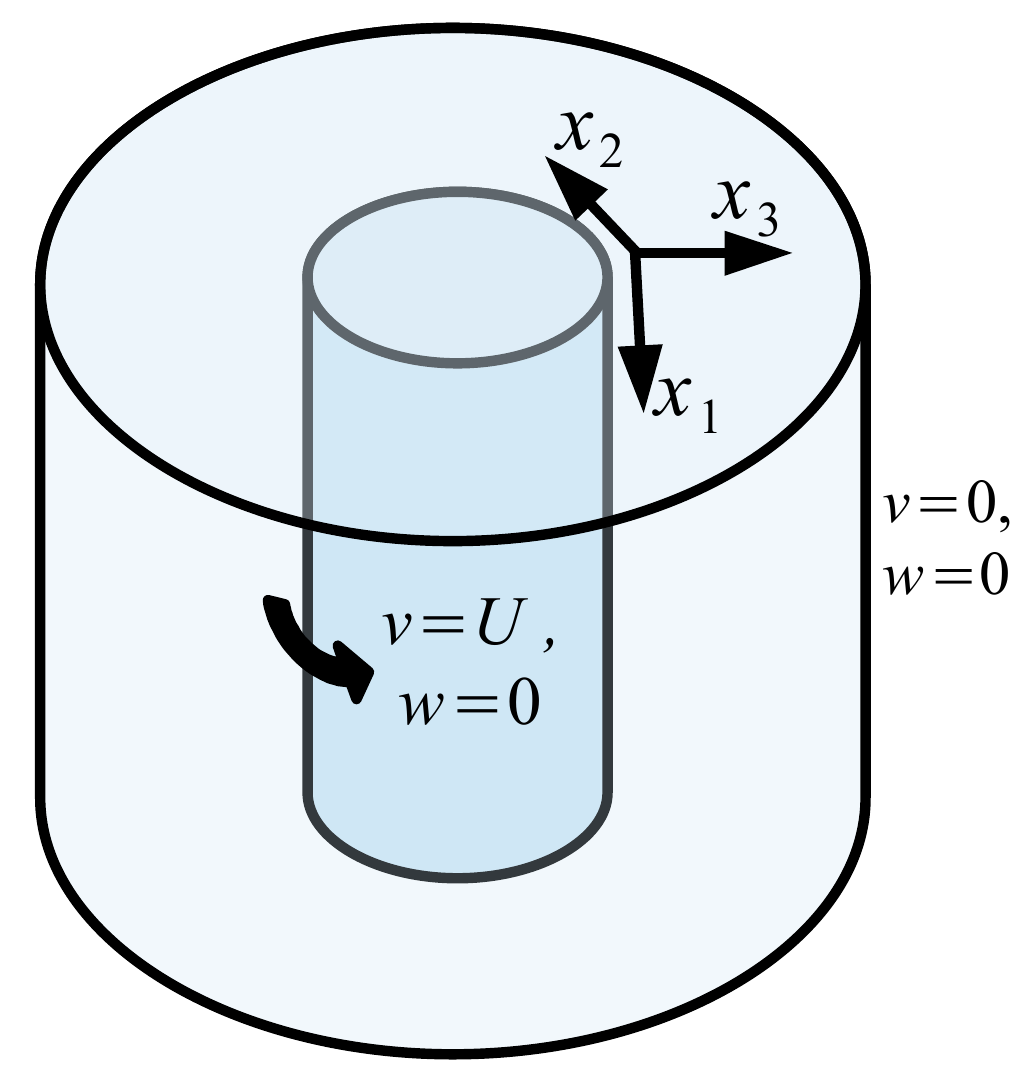}} \\
{\bf (a)} & {\bf (b)}
\end{array}$
\end{center}
\caption{{\bf{(a)}} The geometry, choice of coordinates, and generic boundary conditions for the Rayleigh-B\'enard setup (particular boundary conditions on $u$ depend on the physical setting, i.e. stress-free versus no-slip).  {\bf{(b)}} The same for Taylor-Couette.}
\label{fig:results}
\end{figure}

The dimensional temperature is $\hatt{T}=\hatt{T}(\hatt{x}_1,\hatt{x}_3)$ and it satisfies the boundary conditions 
$\hatt{T}(\hatt{x}_3=0)=T_0+\delta T$ and $\hatt{T}(\hatt{x}_3=d)= T_0$, i.e. the temperature at the bottom plate is 
higher by the fixed amount $\delta T$. 
The equations of motion are
\begin{subequations}
\begin{equation}
\hatt{\partial}_t \hatt{u} + (\hatt{{\bf u}}\cdot\hatt{\nabla}) \hatt{u} + 1/\rho \hatt{\partial}_1 \hatt{p} =  \nu \hatt{\Delta} \hatt{u},
\end{equation}
\begin{equation}
\hatt{\partial}_t \hatt{w} + (\hatt{{\bf u}}\cdot\hatt{\nabla}) \hatt{w} + 1/\rho \hatt{\partial}_3 \hatt{p} = \nu \hatt{\Delta} \hatt{w} + g  \beta \hatt{T}, 
\end{equation}
\begin{equation}
\hatt{\partial}_t \hatt{T} + (\hatt{{\bf u}}\cdot\hatt{\nabla}) \hatt{T} = \kappa \hatt{\Delta} \hatt{T},
\label{eq:T}
\end{equation}
\end{subequations}
combined with incompressibility: $\nabla \cdot {\bf u} = 0$, where $\hatt{\nabla}=(\hatt{\partial}_1, \hatt{\partial}_3)$, 
and $\hatt{\Delta} = \hatt{\partial}_{11}^2+\hatt{\partial}_{33}^2$. The other variables are
the kinematic viscosity $\nu$, the thermal diffusivity $\kappa$, the expansion
coefficient $\beta$ and the gravitational constant $g$.
As mentioned above, $\hatt{x}_3=0$ is the bottom plate and $\hatt{x}_3=d$ the top plate.  In the absence of convection, the
temperature displays a linear profile,  $\hatt{T}_0(\hatt{x}_3)=\delta T\, (1-\hatt{x}_3/d)$.  
This non-convective buoyancy is balanced by the pressure field $\hatt{p}_0(\hatt{x}_3) = \beta g \delta T (\hatt{x}_3 - \hatt{x}_3^2/(2d))$.  We decompose the full temperature field according to $\hatt{T}(\hatt{x_1},\hatt{x_3},\hatt{t})=\hatt{T}_0(\hatt{x}_3) + \hatt{\theta}(\hatt{x}_1,\hatt{x}_3,\hatt{t})$.  
Decomposing the pressure in a similar manner, but using $\hatt{p}$ to now refer to perturbations about the laminar pressure $\hatt{p}_0$, we can derive the
equations for the deviation $\hatt{\theta}$  from the diffusive profile:
\begin{subequations}
\begin{equation}
\hatt{\partial}_t \hatt{u} + (\hatt{{\bf u}}\cdot\nabla) \hatt{u} + \hatt{\partial}_1 \hatt{p} = \nu \hatt{\Delta} \hatt{u},
\end{equation}
\begin{equation}
\hatt{\partial}_t \hatt{w} + (\hatt{{\bf u}}\cdot\nabla) \hatt{w} + \hatt{\partial}_3 p = \nu \hatt{\Delta} \hatt{w} + g  \beta \hatt{\theta},
\end{equation}
\begin{equation}
\hatt{\partial}_t \hatt{\theta} + (\hatt{{\bf u}}\cdot\nabla) \hatt{\theta} - {\delta} \hatt{T}\, \hatt{w} /d  = \kappa \hatt{\Delta} \hatt{\theta},
\end{equation}
\end{subequations}
coupled with incompressibilty for the velocity field.

In RBC, dimensionless variables for temperature, length, and time are based on the temperature
difference $\delta T$, the height $d$ and the thermal diffusivity $\kappa$. Then in the non-dimensional setting, we have the system:
\begin{subequations}
\begin{equation}
\partial_t u + ({\bf u}\cdot\nabla) u + \partial_1 p= \Pra \Delta u,
\end{equation}
\begin{equation}
\partial_t w + ({\bf u}\cdot\nabla) w + \partial_3 p= \Pra \Delta w + \Pra \Ray \theta,
\label{RB_uw}
\end{equation}
\begin{equation}
\partial_t \theta + ({\bf u}\cdot\nabla) \theta = w +  \Delta \theta,
\label{RB_theta}
\end{equation}
\end{subequations}
again coupled with incompressibility of ${\bf u}$, and where the Rayleigh number is given by $\Ray=\frac{g  \beta \delta T\, d^3}{\kappa\nu}$, 
and the Prandtl number $\Pra=\frac{\nu}{\kappa}$.
We now wish to re-scale the 2D RPC system so that it has the same functional form as these three equations.  As we see, the analogy is exact only when $\Pra = 1$, and if we carefully re-scale the azimuthal component of the velocity field so that it appears in the same way as the temperature fluctuations.

\subsection{Quasi 2D Taylor-Couette}
The full Taylor-Couette system consists of three velocity components that are linked by incompressibility. 
In order to split off a temperature-like component, 
we consider a restricted geometry where the fields do not depend on the azimuthal coordinate and we investigate the limit of large radius. In this setup the azimuthal velocity component decouples and can be re-scaled to resemble the temperature fluctuations from RBC.

The radial direction in the TCF-system is the analog of the gravitational direction in 
the RBC-system, so that the $\hatt{x}_3$ direction becomes the radial one. The
neutral direction for the flows is the axial one, which therefore is referred to as the $\hatt{x}_1$ direction, and 
in which we assume the flow to be periodic.
Finally, the azimuthal component becomes the $\hatt{x}_2$ direction. Invariance of the azimuthal position then implies that the velocity field has dependencies
$(\hatt{u}(\hatt{x}_1,\hatt{x}_3), \hatt{v} (\hatt{x}_1,\hatt{x}_3), \hatt{w}(\hatt{x}_1,\hatt{x}_3))$.
In a frame of reference rotating with
frequency $\Omega$ around the $1$-axis the system then becomes
\begin{subequations}
\begin{equation}
\hatt{\partial}_t \hatt{u} + (\hatt{{\bf u}}\cdot\hatt{\nabla}) \hatt{u} + \hatt{\partial}_1 \hatt{p} =  \nu \hatt{\Delta} \hatt{u},
\end{equation}
\begin{equation}
\hatt{\partial}_t \hatt{w} + (\hatt{{\bf u}}\cdot\hatt{\nabla}) \hatt{w} + \hatt{\partial}_3 \hatt{p} = \nu \hatt{\Delta} \hatt{w}  + 2\Omega \hatt{v},
\end{equation}
\begin{equation}
\hatt{\partial}_t \hatt{v} + (\hatt{{\bf u}}\cdot\hatt{\nabla}) \hatt{v} =  -2\Omega \hatt{w} + \nu \hatt{\Delta} \hatt{v},
\label{eq:v_azimuth}
\end{equation}
\end{subequations}
together with incompressibility, $\hatt{\partial}_1 \hatt{u} + \hatt{\partial}_3 \hatt{w} = 0$.

The domain is bounded by two walls, i.e. the inner and outer cylinders, 
which we denote as $\hatt{x}_3=0$ and $\hatt{x}_3=d$, that are parallel to the $\hatt{x}_1-\hatt{x}_3$ plane. 
Between the plates, there is a mean shear, maintained by moving
the walls at constant speed in the $2$-direction.  This yields the boundary conditions 
$\hatt{v}(\hatt{x}_3=0)=U$ and $\hatt{v}(\hatt{x}_3=d)=0$, which gives rise to a linear laminar velocity profile, 
$\hatt{v}(\hatt{x}_3)=U(1- \hatt{x}_3/d)$.  The typical physically motivated boundary condition on the other components of velocity is the no-slip condition meaning that the other components of the velocity field vanish identically at these walls.  In Section \ref{sect:freeslip} we discuss the effect of considering a slippery boundary for $\hatt{v}$ and $\hatt{w}$, a condition that is not as physically relevant but is more conducive to analysis.

As in the case of RBC, we are primarily concerned with deviations $v'$ from the 
linear profile, that is
\beq
\hatt{v}(\hatt{x}_1,\hatt{x}_3,\hatt{t})=U(1- \hatt{x}_3/d) + v'(\hatt{x}_1,\hatt{x}_3,\hatt{t}),  \
\eeq
where $v'$ is the dimensional form of the deviations.
The linear part in $\hatt{v}$ is absorbed in the pressure (compensating the centrifugal forces) as was done for RBC,
so that only the fluctuations remain.
The equation for the azimuthal component then becomes
\beq
\hatt{\partial}_t v' + (\hatt{{\bf u}}\cdot\hatt{\nabla}) v' - (U/d) \hatt{w} =  -2\Omega \hatt{w} + \nu \hatt{\Delta} v'.
\eeq
The contribution from the normal velocity has a prefactor proportional to $(U-2\Omega/ d)$ that
can be absorbed in the normal component $v'$ with the rescaling 
\beq
v'=(U-2\Omega d) \theta\,.
\label{v_theta}
\eeq
Note that this scaling introduces an asymmetry between the velocity components since it 
affects only one and not all three components. It is therefore not
reasonable for the full 3D system. 

The identification of this renormalization of the azimuthal velocity fluctuations to the variable $\theta$ is intentional, as this normalized dependent variable is identified with the temperature fluctuations for the 2D RBC system described above.
With this definition of $\theta$,  the dimensional equation for the normal velocity becomes
\beq
\hatt{\partial}_t \theta + (\hatt{{\bf u}}\cdot\hatt{\nabla}) \theta  =  \frac{1}{d}\hatt{w} + \nu \hatt{\Delta} \theta.
\eeq
and the evolution of $\hatt{u}$ and $\hatt{w}$ in RPC become:
\begin{subequations}
\begin{equation}
\hatt{\partial}_t \hatt{u} + (\hatt{{\bf u}}\cdot\hatt{\nabla}) \hatt{u} + \hatt{\partial}_1 \hatt{p} =  \nu \hatt{\Delta} \hatt{u},
\end{equation}
\begin{equation}
\hatt{\partial}_t \hatt{w} + (\hatt{{\bf u}}\cdot\hatt{\nabla}) \hatt{w} + \hatt{\partial}_3 \hatt{p} = 
\nu \hatt{\Delta} \hatt{w}  + 2\Omega (U/d-2\Omega) \theta.
\end{equation}
\end{subequations}

Now we introduce dimensionless variables for the rest of the system using the height $d$ and the viscosity $\nu$ to generate spatial and temporal scales (and hence velocity as well), which give 
the shear Reynolds number $\ReyS=Ud/\nu$, and the rotation number, $R_\Omega=2\Omega d/U$.
Then the full non-dimensional equations for RPC become
\begin{subequations}
\begin{equation}
\partial_t u + ({\bf u}\cdot\nabla) u + \partial_1 p= \Delta u,
\end{equation}
\begin{equation}
\partial_t w + ({\bf u}\cdot\nabla) w + \partial_3 p= \Delta w + \ReyS^2 R_\Omega (1-R_\Omega) \theta,
\end{equation}
\begin{equation}
\partial_t \theta + ({\bf u}\cdot\nabla) \theta = w + \Delta \theta,
\end{equation}
\end{subequations}
which is formally identical to the RBC case identifying
$\Pr=1$ and $\Ray=\ReyS^2 R_\Omega (1-R_\Omega).$
This derivation did not use the boundary conditions on the velocity field so that it is valid for both rigid and free slip boundary conditions or any mixture thereof.

The exact relationships between the RPC and RBC systems considered here are summarized in table \ref{table:problems}.

\begin{table}
\caption{Exact relations between the azimuthally independent, large radius TCF system, and 2D RB.}
\centerline{\begin{tabular}{ccc}  \hline\hline \label{table:problems}
~~ & \textbf{Rayleigh-B\'enard} & \textbf{Taylor Couette} \\  \hline
\textbf{corresponding velocities} & $u$,~$w$ & $u$,~$w$ \\
\textbf{temperature and azimuthal velocity
} & $\theta$ & $\left(
{U}/{d} - 2\Omega\right)v$ \\
\textbf{Prandtl number} & $\Pra$ & $1$ \\
\textbf{Rayleigh, Reynolds, and rotation numbers} & $\Ray$ & $\ReyS^2 R_\Omega(1-R_\Omega)$ \\ 
\textbf{Nusselt numbers} &$\Nus_T-1$ &$ 
({\Nus_S-1})/({1-R_\Omega})$
\end{tabular}}
\end{table}

\section{Consequences of the relation between RPC and RBC}
\subsection{Heat and momentum transport}

\begin{wrapfigure}{r}{.5\textwidth}
	\centering
{\includegraphics[width=0.45\textwidth]{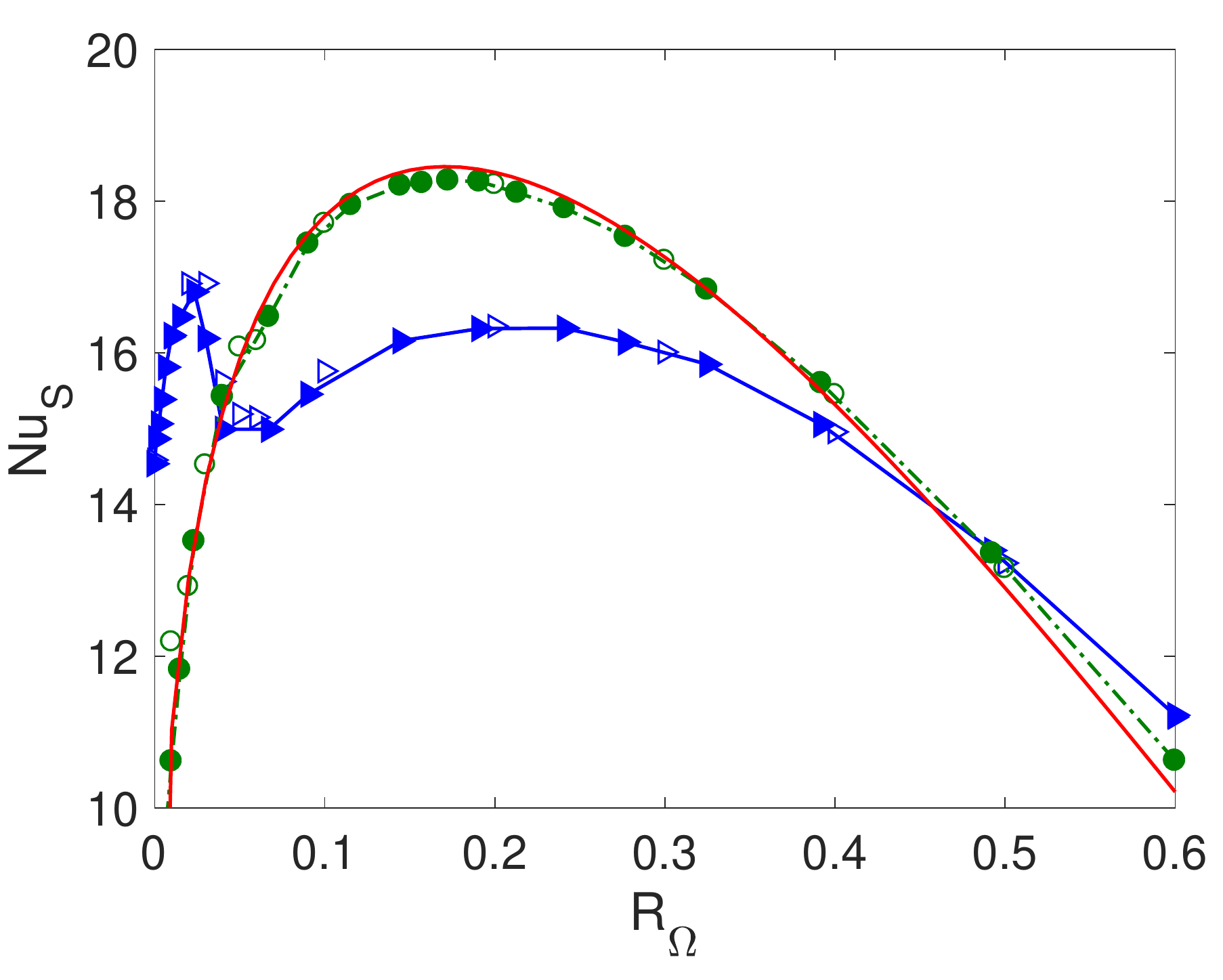}}
\caption{Nusselt number $\Nus_S$ for TCF and RPC versus rotation number $R_\Omega$ for $\ReyS = 2 \times 10^4$, redrawn from Figure 8f of \cite{Brauckmann:2016ji}.  The full symbols are for TCF at radius ratio $\eta=0.99$, the open symbols are for RPC.  The blue data are for the full three-dimensional flow; they have a broad maximum at $R_\Omega \approx 0.22$, and a narrow one for smaller $R_\Omega$ of a different origin (see \cite{Brauckmann:2016ji}).  The green data are obtained from the azimuthally invariant 2D part of the flow.  The continuous red line is a fit of the 2D part to expression (\ref{eq:NuS_approx}), with optimal parameters $\alpha=0.26$ and $c=37$ and a maximum near $R_\Omega \approx 0.18$, very close to the maximum obtained from (\ref{Romegamax}).}\label{fig:data_compare}
\vspace{-.5cm}
\end{wrapfigure}

The first set of consequences we discuss here stem from the heat and angular momentum transport,
and will be 
valid for all boundary conditions on the velocity fields $u$ and $w$. 
In RBC, the temperature difference
drives convection which enhances the transport between the plates. The Nusselt number
measures the heat transport in units of the diffusive heat transport, and is computed as $\Nus_T = 1 + \overline{w \theta}$,
where $\overline{\cdot}$ refers to an average in time, and across the neutral direction $x_1$. This quantity
is independent of $x_3$.

For RPC, the azimuthal momentum transport is derived by averaging the dimensional equation
for the azimuthal velocity 
in the $\hatt{x}_1$ direction: $\hatt{\partial}_3  \overline{\hatt{w}\hatt{v}} = 2\Omega \overline{\hatt{w}} + \nu \hatt{\partial}_{33}^2 \overline{\hatt{v}}$.
The average over $\hatt{w}$ vanishes by incompressibility: physically, such a non-zero average would mean that there is a 
non-zero mean flux of fluid in the 3-direction, which is not possible. This can also be shown by taking the area average of the incompressibility condition: $\hatt{\partial}_3 \overline{\hatt{w}} = -\hatt{\partial}_1\overline{\hatt{u}}$,
by periodicity in $\hatt{x}_1$.
Since this quantity must vanish at the walls (for all relevant boundary conditions), 
it vanishes everywhere.  Integrating once in $\hatt{x}_3$ then gives the momentum current
\beq
J = \overline{ \hatt{w}\hatt{v}} - \nu \hatt{\partial}_3\overline{\hatt{v}} = -\nu \hatt{\partial}_3\left.\overline{\hatt{v}}\right|_{\hatt{x}_3=0},
\eeq
which is independent of the position $\hatt{x}_3$ between the plates. 
Using the laminar profile and (\ref{v_theta}) for the azimuthal velocity and
dividing $J$ by the linear viscous drag $\nu U/d$, 
one arrives at the equivalent of the Nusselt number in RPC for the non-dimensional variables
\begin{align}
\Nus_S &= 1 + \left(1-\frac{2\Omega d}{U}\right) \overline{w\theta} = 1+\left(1-R_\Omega\right) \overline{w\theta},\\
&\Rightarrow \Nus_S(\ReyS, R_\Omega) -1 = (1-R_\Omega) (\Nus_T (\Ray)-1).
\end{align}
Thus, while the relation between the parameters $\Ray$ in RBC and $\ReyS$ and $\Rom$ in RPC
is symmetric under the exchange $\Rom$ to $1-\Rom$, the relation between heat and 
momentum transport is not.  Remarkably, this connection predicts that $\Nus_S$ approaches the laminar value $\Nus_S=1$ for $\Rom$ approaching $1$, for any value of $\ReyS$. 

The transport of momentum as a function of rotation number has been studied 
for  different parameter values in both TCF and RPC.  Assuming that the Nusselt number 
in RBC follows a scaling law $\Nus_T \sim c \Ray^\alpha$ with an as yet undetermined
exponent $\alpha$ and some constant $c$, we can obtain a scaling for 
$\Nus_S$ in terms of both $\ReyS$ and $R_\Omega$
and determine the maximal $\Rom$.  First, we observe that
\begin{align*}
 \Nus_S -1 &\sim c \ReyS^{2\alpha}R_\Omega^\alpha(1-R_\Omega)^{1+\alpha} - 1 + R_\Omega\\
\Rightarrow \Nus_S &\sim R_\Omega + c \ReyS^{2\alpha} R_\Omega^\alpha(1-R_\Omega)^{1+\alpha}.
\end{align*}
For large $\ReyS$, the first part can be neglected and the relation reduces to
\begin{equation}\label{eq:NuS_approx}
\Nus_S \sim c' R_\Omega^\alpha (1-R_\Omega)^{1+\alpha},
\end{equation}
away from $R_\Omega = 0$ or $R_\Omega = 1$.
Maximizing this transport over the rotation rate $R_\Omega$ (for $\ReyS \gg 1$) leads to 
\begin{equation}
\mbox{R}_{\Omega,m} \sim \frac{\alpha}{1+2\alpha}\,.
\label{Romegamax}
\end{equation}
In one conjectured asymptotic regime of thermal convection \citep{Sp1963} it is expected that $\alpha={1}/{2}$ so that the maximal momentum transport would occur for $\mbox{R}_{\Omega,m} \sim {1}/{4}$.  The Reynolds numbers in numerical simulations and even in experiments are not high enough to reach this regime, and one resorts to Reynolds number dependent local scaling exponents.  For RPC and $\Rey_S = 2\times 10^4$,  \cite{Salewski:2015co} find a  maximum near $\mbox{R}_{\Omega,m}\approx 0.2$.  The transition from TCF to RPC is discussed in \cite{Brauckmann:2016ji}, where it is shown that the maximum again appears near $\mbox{R}_{\Omega,m}\approx 0.2$ when the ratio of the radius of the inner cylinder to that of the outer cylinder is $0.99$.  This observation is further cemented by the recent numerical and experimental results reported in \cite{ezeta2020double}.  Both sets of data (for TCF and RPC) are shown in Figure \ref{fig:data_compare} (blue symbols).  This agreement with observations is remarkable because the derivation here is valid only for the 2D setting with a rescaling of $v$ that violates the natural scaling of the velocity fields, whereas the reported maxima of $\Rom \sim 0.2$ comes from numerical studies of the full 3D flow.  However, since rotation reduces the transverse components and enhances the azimuthally invariant parts, \cite{Brauckmann:2016ji} also isolated the contribution of the azimuthally invariant 2D component of the flow, shown in Figure \ref{fig:data_compare} as the green data points.  They show a maximum near $R_\Omega \approx 0.18$.  Moreover, a fit of the 2D contributions to the functional form (\ref{eq:NuS_approx}) with $\alpha=0.26$ approximates the data very well over essentially the entire range $0 < R_\Omega < 0.6$ for which data is available.  This confirms that the 2D analysis presented here can also explain features of the three-dimensional flow.


\subsection{Mean profiles and the maximum principle}
In turbulent RBC flows one expects the temperature to be well mixed in the interior, so that
the profile consists, to a good approximation, of steep boundary layers near the top
and bottom plates, and a region of constant temperature in the middle.  
We anticipate the same behavior will hold for the azimuthal velocity $v$ relative to the $\hatt{x}_3$ direction.
Moreover, since equation (\ref{eq:T}) is an explicit advection-diffusion equation,
the temperature satisfies a maximum principle, i.e. $\hatt{T}$ is bounded by its values
at the walls, 
\beq
T_0\le \hatt{T}\le T_0+\delta T,
\eeq 
at any point in the volume. This arises because near a maximum the first derivative vanishes and the second is negative, so diffusion will act to reduce it. For RBC this indicates that there heat cannot pile up at the walls beyond that supplied by the boundary condition. 
Translated to RPC, the corresponding statement will address the relation between the downstream
velocity and its value at the walls. A velocity field that lies outside the values provided by the boundary conditions is refered to in the literature as a backflow event. Existence of such backflow events has been a matter
of controversy which was settled with the explicit demonstration of such events by \citet{Lenaers:2012cz}.

To see how the maximum principle applies to RPC, we let $\hatt{v}=-2 \Omega \hatt{x}_3 + \tilde v$, 
so that $\tilde v$ satisfies the advection diffusion equation $\hatt{\partial}_t \tilde v + (\hatt{{\bf u}}\cdot\hatt{\nabla}) \tilde v =  \nu\hatt{\Delta} \tilde v$.  This perturbative velocity field $\tilde v$ satisfies a maximum principle, i.e. 
its extreme values occur only at $\hatt{x}_3=0$ or $\hatt{x}_3 = d$.  
This can be stated succinctly as $2\Omega d \leq \tilde v \leq U$, or $0\leq \hatt{v} \leq U$ for the original azimuthal velocity component (assuming a positive $U$).  This implies that in this setting there is no backflow, 
i.e. the azimuthal velocity has a constant sign, and events like the ones
observed by \cite{Lenaers:2012cz} can not occur in 2D RPC.  It would be of significant interest to determine if such events can occur for 3D RPC or even for the full TCF system.  If so, one must question whether these events are present only for full three-dimensional flows or if the current restriction in RPC is unique.

\subsection{Free slip boundary conditions}
\label{sect:freeslip}
Although the no-slip boundary condition is clearly the physically motivated choice for TCF, and hence for RPC as well, there is still some interest in considering the free-slip condition although there is less immediate physical motivation for this.
Indeed, \cite{Ra1916} invoked free-slip conditions for mathematical convenience:
\begin{quote}
\emph{...for a further condition we should probably prefer $dw/dz = 0$ }[no-slip]\emph{, corresponding to a fixed solid wall.  But this entails much complication, and we may content ourselves with the supposition $d^2w/dz^2=0$} [free-slip]...
\end{quote}
Thus, in the interest of reducing such \emph{complication} due to the no-slip condition, we will consider stress-free (free-slip) boundary conditions on the plates for $u$ and $w$.  As mentioned previously this is presented in the form $w=0$ and $\partial_3 u = 0$ at $x_3=0$ and $1$.  Incompressibility then implies that the vorticity in the $x_2$ direction defined by $\omega_2  = \partial_1w-\partial_3u$ vanishes at these boundaries as well.
This leads to an enstrophy balance, where the enstrophy is defined as the square of the L$^2$ 
norm of $\omega_2$, i.e. $\|\omega_2\|_2^2$.  This restriction which we consider in this Section only, does not reflect on any other aspect of the system as described above.

As shown by \cite{WhDo2011a} this enstrophy balance, coupled with a uniform bound on $w(x_3)$ near the boundary 
and a piece-wise linear, monotonic temperature profile, will yield a bound on the Nusselt number in RB 
of the form $\Nus_T \leq c \Ray^{5/12}$.  Translated into the RPC system, this becomes
\beq
\Nus_S \lesssim R_\Omega + \ReyS^{5/6}(1-R_\Omega)^{17/12}R_\Omega^{5/12},
\eeq
which will have a maximum as $\ReyS \rightarrow \infty$ for $\mbox{R}_{\Omega,m} = \frac{5}{22} \sim 0.227$
by equation (\ref{Romegamax}).
This is the only setting to date for which the scaling of $\Nus_S$ is sublinear with respect to $\ReyS$, 
deviating from the anticipated  $\ReyS^{1}$ scaling.  Remarkably this scaling affects the $\ReyS$-dependence, but it has little influence on the
location of the maxima in $\Rom$.

\section{Conclusions and discussion}
We have analyzed some consequences of an exact relationship between 2D RBC and RPC, the limit of 2D TCF flow for large radii. For these settings the two problems can be mapped onto each other via an identification of corresponding fields and  a change of variables as defined in table \ref{table:problems}.  This converts the well-known analogy between these systems to an exact formulation in 2D.  Comparison of these results to previous numerical and experimental observations indicate that some of the results apply to the full three dimensional situation.
For instance, the additional factor in the relation between heat and momentum transport gives an asymmetry in the
location of the maximum as a function of $\Rom$, which agrees well with observations 
on fully three-dimensional flows.

Restricting to the 2D setting allows for the absence of backflows due to 
a maximum principle, and, for the specific choice of free slip boundary conditions,
the results on the sub-linear scaling of torque with shear Reynolds number. In both cases
it should be interesting to explore further how higher dimensions and modifications of 
boundary conditions can cause deviations from these 2D relations.  More generally, the 
differences in the equations of motion between the 2D and 3D cases may point to other 
observables in which fully 3D RB convection and RPC differ.

Declaration of Interests. The authors report no conflict of interest

\section*{Acknowledgements}
Prior to completion of this paper the first author, our dear friend Bruno Eckhardt, died rather suddenly. Bruno inspired the research reported here and contributed the central ideas to the study. This is not the place to describe Bruno's lasting contributions to the study of nonlinear dynamics and turbulence; suffice it to say that his loss has been and will be deeply felt not only by his closest collaborators, but by all who have an interest in the field. Our effort to complete this paper is dedicated to his memory.  We also thank Hannes Brauckmann for providing the data necessary to produce Figure \ref{fig:data_compare}. This work was supported in part by the US National Science Foundation via awards DMS-1515161 and DMS-1813003, and the Simons Foundation through award number 586788. This work was initiated at the Institute for Pure \& Applied Mathematics \emph{Mathematics of Turbulence} program during the fall of 2014, and we appreciate the persisting influence of that program.


\end{document}